\documentclass[letterpaper]{article}
\usepackage{aaai}
\usepackage{times}
\usepackage{helvet}
\usepackage{courier}
\usepackage{epsfig}
\usepackage{subfigure}
\usepackage{calc}
\usepackage{amssymb}
\usepackage{amstext}
\usepackage{amsmath}
\usepackage{multicol}
\usepackage{pslatex}
\usepackage{enumerate}
\usepackage{fancyhdr}
\usepackage[small]{caption}
\usepackage{color}

\title{\uppercase{Characterising Through Erasing}\\
 \fontsize{13}{15}\selectfont \textit{A Theoretical Framework for Representing Documents Inspired by Quantum Theory}}
 \author{A. F. Huertas-Rosero \and L. A. Azzopardi \and C. J. van Rijsbergen\\
 Dept. of Computing Science, University of Glasgow, \\ Glasgow, United Kingdom\\
  \{alvaro, leif, keith\}@dcs.gla.ac.uk
  }
\begin{document}

\newcommand{\todo}[1]{\textcolor{red}{#1}}
 \maketitle

\begin{abstract}
\fontsize{9}{11}\selectfont
The problem of representing text documents within an Information Retrieval system is formulated as an analogy to the problem of representing the quantum states of a physical system. Lexical measurements of text are proposed as a way of representing documents which are akin to physical measurements on quantum states. Consequently, the representation of the text is only known after measurements have been made, and because the process of measuring may destroy parts of the text, the document is characterised through erasure. The mathematical foundations of such a quantum representation of text are provided in this position paper as a starting point for indexing and retrieval within a ``quantum like'' Information Retrieval system.
\end{abstract}

\section{Introduction}
 The problem of indexing, i.e. generating compact and informative representations of documents, is an important issue in Information Retrieval (IR). For text documents, the most successful representations have been based on the occurrence of terms in documents. Either their presence or absence, or some statistical information about the term's occurrence in the document. Consequently, a document is represented as a array of terms, and assumed to be fixed or static in nature. These representations are used in standard IR models such as the Boolean model, Binary Independence Model (BIM), Vector Space Model, Language Model, etc~\cite{RepresentationInIR,ponte-croft98,salton68vsm} where the representation employed tends to be dictated by the model. For example, both the Boolean model and BIM expect a binary representation, whereas the Language Model expects a probability distribution over the vocabulary.

 In this work, a different approach is taken, where instead of focusing directly on building an IR model, the focus is put on devising an underlying representation of documents, which is inspired by Quantum Theory (QT).  Such a representation should be suitable for being used by an IR system.

 An important part of physics deals with the problem of representing in the state of a system, the information an observer can obtain from a set of measurements. QT provides a solution in which measurements on a quantum system can be obtained to provide a representation of the state of the system. This theory is based on the science of natural objects (i.e. photon, electrons, etc). However, IR is a science of artificial objects (i.e. text / documents)~\cite{QuantumIR}. Consequently, it is necessary to explain how QT can be applied in the context of IR.
 
 Documents can be thought of as states of a physical system, and their features (such as terms), can be viewed as physical observables to be measured in such system. If a suitable definition of the measurements to be performed on documents is used, then the powerful theoretical machinery of QT can be engaged to represent and use the information obtained. The main contribution of this position paper is to define suitable lexical measurements which can be performed on text which will form the basis for a document representation scheme.

Historically, the most successful methods for automatic indexing of text documents have been based mainly on the statistical analysis of the occurrence of terms in documents~\cite{ShallowData}. It is reasonable, therefore, to propose measurements which are based on the features related to the frequency of occurrence of terms in text documents. These will be referred to as lexical measurements. 

In the next section, lexical measurements on text documents are proposed and defined, and it is shown how these measurements reflect the properties of ideal quantum measurements. Then, operations between the measurements are defined, which enable different relationships to be captured. The proposed measurements are then discussed and directions for further work outlined. 
 
\section{Lexical measurements on Textual Documents}\label{LexicalMeasurements}

In a physical system, the state of the system is defined by the probabilities of the possible outcomes of measurements performed on that system. However, the state of a quantum system can only have some of the measurement outcomes determined, not all of them.  For example, there is an impossibility of determining both position and velocity of an electron (Heisenberg indeterminacy principle): only one of the two properties can be determined with certainty, while the other becomes uncertain when the first is determined.

For some pairs of measurements, the value of the corresponding observables will not depend on the order in which the measurements are performed.  In that case, measurements are \textit{compatible}.   Other measurements, however, interfere with each other, in such a way that the obtained outcomes depend on the order in which measurements are performed.  These are \textit{incompatible} measurements.

A maximal set of compatible measurements can be performed in order to determine the state to a maximum extent, but measurements that are incompatible will not have their outcomes determined. This maximal set chosen by the experimenter can be thought as an experimental context, and the system will have less information about the outcomes of other sets of measurements that are incompatible to the chosen ones.

The problem of using information from measurements to represent the state of a system can be formulated as a very general representation problem.    It is possible to adopt the view of lexical measurements as physical measurements on a system, and use a sophisticated representation scheme borrowed from physics.   The involved measurements must then be defined in such a way that they are akin to the properties of physical measurements, described above. The proposed lexical measurements are based on measuring the co-occurrence of terms within documents to act like a measurements on a quantum system.   Counting would be viewed as a projective operation on text documents, via certain transformations of the document that are defined in the next section.

\subsection{Selective Erasers}
The proposed approach is based on the definition of certain transformations that can be applied to text documents.    These transformations will be called \textbf{Selective Erasers} and are denoted by $E(t,w)$, where $t$ is a chosen central term, and $w$ is the number of preserved terms on either side of the occurrence of $t$.   Applying a Selective Eraser amounts to erasing every term in the document not falling within a window of text (a sequence of terms in the document) centred in an occurrence of $t$, which includes $w$ tokens to the left and $w$ tokens to the right (see figure 1).  Thus, the total size of the window is $2\times w+1$ tokens. 
 \begin{figure}
    \caption{The Selective Eraser}
    \label{Eraser}
    \begin{center}\includegraphics{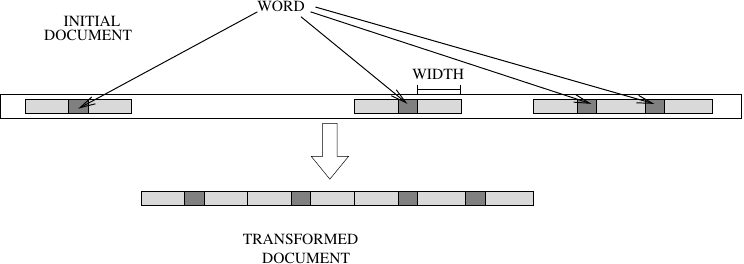}\end{center}
    {\small The big white box represents a document, the gray areas represent the chosen window, and the dark gray squares represent the occurrences of the chosen term in the middle of each window.}
   \end{figure}
 We can define a transformation $E(t,w)$ that converts document $D$ into document $D'$ with some erased tokens, such that $E(t,w)D = D'$.

 In the following subsection, the ``quantum like'' properties of erasers are described along with the operations that can be performed on them, which extend the possibility of using them beyond simple co-occurrence measures, and form the basis of a ``quantum like'' IR system.

\subsection{Properties of Selective Erasers}
 According to \citeauthor{QuantumMeasurements} (\citeyear{QuantumMeasurements}), ideal quantum measurements need to satisfied three important properties: (1) idempotency (projection postulate), (2) an ordered structure, and (3) the possibility of being non-commutative.   Given the definition of the Selective Eraser, each property is fulfilled as described below:
  \begin{enumerate}
   \item They are idempotent: applying them any number of times is the same as applying them once.  For example: let document $D=$``to be or not to be, that is the question''.   If we apply $E(is,2)$ to $D$, we are left with $D'=$``be, that is the question''.  If we apply it again, it will not perform further deletions, because all terms are within the window already: $E(is,2)D = E(is,2)\left[ E(is,2)D\right] $.
   
   \item They have \textbf{order relations} (see figure 2). If applying $E_1$ and then $E_2$ gives the same result as applying $E_1$, then we can say $E_2 \leqslant E_1$.  With the same example, we can compare $E(is,2)$ and $E(is,3)$.  If a term is erased by $E(is,3)$ it will also be erased by $E(is,2)$, but not necessarily the other way around.  In our example with $D$, both would erase ``To be or not'', but only $E(is,2)$ would erase the second ``to''.   So, we could say that $E(is,3) \geqslant E(is,2)$ because $E(is,3)$ will always leave unchanged the same terms as $E(is,2)$, and possibly other terms.   The mathematical definition of the order relation is:
   \begin{equation}\label{OrderRelation}
     E_1 \geqslant E_2 \iff \forall D_i: E_2 \left[ E_1 D_i \right] = E_2 D_i
   \end{equation}
   When they do not have an order relation, we can say they are incompatible, and represent that relation with the symbol $\ncong$.
   {\small \begin{equation}
     E_1 \ncong E_2 \Rightarrow
         \lnot ( E_1 \geqslant E_2 ) \land \lnot( E_1 \leqslant E_2 )
   \end{equation}}

   \begin{figure}[th]
    \caption{Order relations between compatible erasers}
    \label{Order}
    \begin{center}\includegraphics{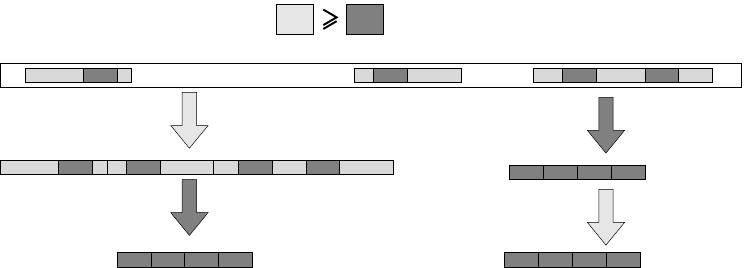}\end{center}
    {\small Here the lighter gray areas represent one eraser, and the dark areas another.   These two erasers are said to be compatible because the result is the same in any order: they commute.    They also show an order relation: one of them includes the other because it preserves the same parts of the document, plus others.}
   \end{figure}

 \item They do not always commute.  When some terms in a document are erased by both projectors $E_1$ and $E_2$, and some occurrences of the central term $t_i$ of one is amongst them, it is easy to see that applying the erasers in a different order produces a different result (see figure 3). 

  This is similar to the situation we find with measurements in QT: there are particle-like properties, such as position, that are incompatible with wave-like properties, such as wavelength (closely related to velocity).   Measuring a particle-like property will always erase part of the information about wave-like properties, and the other way around, so the result is different when making the two measurements in two different orders.
  \end{enumerate}

\subsection{Operations with Selective Erasers}
An eraser can be thought of as a selection of terms fulfilling a certain proposition, like ``the term is less than $w$ terms apart from an occurrence of term $t$''.   If the proposition is false, the term is erased, while if the proposition is true, the term is preserved.   Moreover, we can define composite transformations made with erasers in a number of ways.   They can be noted as proposition themselves, and it is possible to operate on them using the usual logical operations, like ``not'' ($\lnot$) ``or'' ($\lor$) and ``and'' ($\land$).   Three such composite transformations that can be defined are:
  \begin{enumerate}
   \item \textbf{The complement}: $\lnot E$ erases every term that is \textit{not} erased by $E$ from the document
   \item \textbf{The join} $E_1 \lor E_2$ erases all the terms that would be erased by both of the erasers
   \item \textbf{The meet} $E_1 \land E_2$ erases the terms that would have been erased by any of the erasers
  \end{enumerate}
   \begin{figure}[bh]
    \caption{Operations between erasers}
    \label{Operations}
    \begin{center}\includegraphics{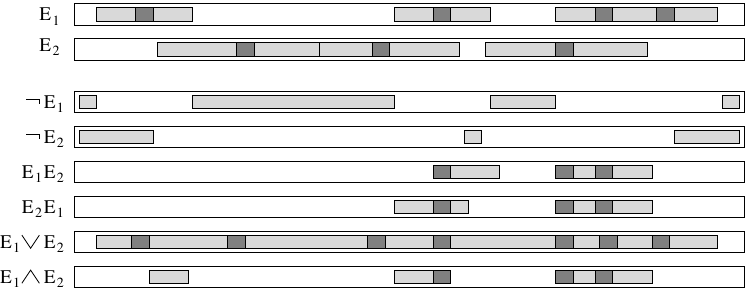}\end{center}
    {\small The big white boxes are the same document, and the dark squares are the occurrences of the chosen terms.   The gray areas are the parts of the document preserved for each transformation.}
   \end{figure}

   It is easy to verify that some order relations always hold for these composite transformations:
   \begin{align}
    (E_1 \lor E_2) \geqslant &E_1 \geqslant (E_1 \land E_2) \\
    (\lnot E_1 \geqslant \lnot E_2) &\iff (E_1 \leqslant E_2)
   \end{align}

  Some order relations arise from the logical characteristics of the propositions that define the transformations.    Let proposition $P_1$ define transformation $T_1$ and proposition $P_2$ define transformation $T_2$.    If $P_1$ implies $P_2$  ($P_1 \Rightarrow P_2$) then we can infer a order relation between the transformations defined by those propositions: $T_1\leqslant T_2$.   As all the terms fulfilling $T_1$ fulfil also $T_2$, then $T_2$ will leave the same terms or more than transformation $T_1$.

 Other order relations between the transformations are not determined by the logical structure of the propositions, but contingent on the choice of documents.     They will hold for some documents, but not for others. 

 The simplest Selective Erasers are those which erase everything but the occurrence of a term.   According to the definition, they would be referred to as $E(t,0)$.    They will be represented by $1$-dimensional projectors.   If such Selective Erasers are applied to each term in the vocabulary then each projector will be orthogonal to one another, because if we apply one to the document, the result of applying another will erase the remainder:
   \begin{equation}
    E(t_1,0)E(t_2,0) = 0 \iff t_1 \neq t_2
   \end{equation}
 
 The application of an eraser $E(t_i,0)$ on $D$ will produce a transformed document, containing only the occurrences of $t_i$. Using a counting operation on the transformed document, the number of times $t_i$ occurs can be obtained. If an eraser for each term is applied independently on $D$, then the term frequency of each term can be obtained. This will then result in a standard bag-of-words representation of $D$.  For instance,
  $
   N(t_i,D) = |E(t_i,0)D|
  $ where $N(t_i,d)$ the number of times $t_i$ occurs in $D$, and $|.|$ is the counting operation which returns the number of tokens in the transformed document.

 The task of determining co-occurrence of terms in a window~\cite{CoOccurrenceBruza}, can also be expressed in terms of Selective Erasers.  A co-occurrence measurement of terms $t_i$ and $t_j$ within a window of length $w$ can be performed in a similar way, where the number of times  $t_i$ occurs in the vicinity of $t_j$ defined by a window of width $w$, in a document $D$ can be defined by  
  $
   N(t_i,t_j,w,D) = |E(t_j,0)\left[ E(t_i,w)D\right]|
  $.  First, a wide-window Selective Eraser $E(t_i,w)$ is applied to $D$, then, a narrow window eraser $E(t_j,0)$ is applied, and then words are counted in the resulting document.

   Higher order erasers ($w>0$), will capture semantic relations between terms and will also be reflected in the order relations.    For example, for some documents a Selective Eraser centred in the term ``George'' with some width $w$ will hold a relation with those centred in the term ``Bush'' with width $w-1$, because the two terms appear together:
  \begin{equation}\label{GeorgeBush}
   E(George,w)\geqslant E(Bush,w-1)
  \end{equation}
  For other documents, the same would hold for ``Kate'' and ``Bush'':
  \begin{equation}\label{KateBush}
   E(Kate,w)\geqslant E(Bush,w-1)
  \end{equation}

  These relations can be used to define different subsets of documents (clusters): we could define the class of documents where (\ref{GeorgeBush}) holds, and the class of documents where (\ref{KateBush}) holds.   While this example is trivial, when bigger windows are involved, the representation can include more complex particularities in the use of the terms.   In future work, we hope to explore the potential uses of this idea in a clustering scheme.

\section{Probabilities}
  Erasers can be seen as \textit{a proposition about a certain word} (for example: term $t_1$ is in the neighbourhood of term $t_2$) that can be fulfilled or not by any token in a document (like being in the neighbourhood of an occurrence of a certain term).   As such, they can be given a truth value for every token in a document, and such values are logically related for different erasers in the ways explained above.  But it is also natural to assign them probabilities, and this can be done in a very simple way by Gleason's theorem~\cite{GleasonsTheorem}.   For a given state of affairs represented by $\rho$, a probability measure can be defined for erasers in the following way:
  \begin{equation}\label{Probability1}
   P(E) = Trace(\Pi_E \rho)
  \end{equation}
  where $\rho$ is a density operator representing the preparation of the system: it can be a representation of a single document, or a representation of a collection consisting of several or all the documents.  To assign a meaning to this probability, it is important to note that it refers to \textit{any token} in a document.   We could say that it is \textbf{the probability of any token in the document, picked at random, which is left unerased by the transformation (eraser) $E$}.   Beyond this frequentist interpretation of the obtained probability, it is possible to follow a Bayesian interpretation of quantum probability \cite{BayesianQT} and define conditional probabilities that reflect the logical structure inherent in these transformations.   For example, we can define:
  \begin{itemize}
   \item  Since (\ref{Probability1}) can be thought of as the \textbf{probability of any token in the system to be unerased by transformation $E$, given a preparation of the system in state $D$} represented by density operator $\rho_D$.
   \begin{equation}\label{Probability2}
    P(E\vert D) = Trace(\Pi_E \rho_D)
   \end{equation}
    \item We can also define the probability of a token not to be erased by eraser $E_2$, given a preparation in document D and a previous application of eraser $E_1$:
   \begin{equation}\label{Probability2}
    P(E_2 \vert E_1 D) = Trace\left(\Pi_{E_2} (\Pi_{E_1} \rho_D \Pi_{E_1})\right)
   \end{equation}
    \item It is even possible to define the probability of an implication:
  \begin{equation}\label{Probability3}
   P(E_1 > E_2 \ \vert D) = \frac{Trace\left(\Pi_{E_2} (\Pi_{E_1} \rho_D \Pi_{E_1})\right)}{Trace(\Pi_{E_1} \rho_D)}
  \end{equation}
  \end{itemize}
  All these probabilities are computed from the representations of the documents, collections, and erasers. What is their relation to lexical, experimental quantities? The relation is, indeed, simple.   We can define, for lexical measurements in one document, a fraction that will behave as a probability:
  \begin{equation}
   F(ED) = \frac{|ED|}{|D|}
  \end{equation}
  where $|ED|$ is the number of tokens in the document after applying $E$, and $|D|$ is the number of tokens in the initial document.   Probabilities, as we defined them, can be simply equated to these fractions:
  \begin{equation}
   P(E\vert D) = F(ED)
  \end{equation}
  Mathematical representations for erasers and document can be derived from measured fractions $F(ED)$ choosing them as to exactly, or approximately, reproduce these numbers with the traces of their products.   A scheme similar to this has been proposed by Mana~(\citeyear{StatesAndMeasurements}) for probabilistic data analysis, but in a more general context.

\section{Conclusions}
  In this paper we have proposed an approach for the representation of documents based on the analogy between lexical measurements on documents and measurements on physical systems.   This approach allows us to represent not only lexical features that are used in traditional methods (i.e. bag-of-words), but also to include more detailed characteristics of the use of words, like co-occurrence.   However, the approach extends beyond such standard interpretations, and provides order relations between propositions about the relative positions of words. This provides a novel way in which to interpret lexical relations that would not be otherwise possible without the application of this quantum analogy.

 In the future, we hope to develop practical IR applications based on Selective Erasers. To this aim, we will explore two main directions: (1) using order relations of Selective Erasers as a way to define clusters of documents, and (2) formulating an indexing scheme based on a density operator representation of documents, that allows the use of the rich mathematical structure of Hilbert Spaces to encode semantic information about documents.

\textbf{Acknowledgements} We would like to thank Guido Zuccon for his valuable input and suggestions.  This work was sponsored by the European Comission under the contract FP6-027026 K-Space and Foundation for the Future of Colombia COLFUTURO.

\bibliographystyle{aaai}
 
\end{document}